\documentclass[aps,prl,twocolumn,showpacs]{revtex4}
\usepackage{amsfonts}
\usepackage{amsmath}
\begin{document}
\title{Detection of entanglement and Bell's inequality violation}
\author{Kai Chen and Ling-An Wu}
\affiliation{Laboratory of Optical Physics, Institute of Physics,
Chinese Academy of Sciences, Beijing 100080, P.R. China}
\begin{abstract}
We propose a new method for detecting entanglement of two qubits
and discuss its relation with the Clauser-Horne-Shimony-Holt
(CHSH) Bell inequality. Without the need for full quantum
tomography for the density matrix we can experimentally detect the
entanglement by measuring less than 9 local observables for any
given state. We show that this test is stronger than the CHSH-Bell
inequality and also gives an estimation for the degree of
entanglement. If prior knowledge is available we can further
greatly reduce the number of required local observables. The test
is convenient and feasible with present experimental technology.
\end{abstract}
\pacs{03.67.-a, 03.65.Ud, 03.65.Ta}
\date{\today}
\maketitle

Since the well-known debate of Einstein, Podolsky and Rosen
\cite{EPR35} with Schr\"{o}dinger \cite{Sch35} about the
completeness of quantum mechanics, entangled states have intrigued
physicists for decades. In particular, in recent years entangled
states have become the key ingredient in the rapidly expanding
field of quantum information science, with remarkable prospective
applications such as quantum teleportation, quantum cryptography,
quantum  dense coding and parallel computation
\cite{pre98,nielsen,zeilinger}. However, the intrinsic nature of
entanglement is by no means fully understood and the theory is far
from complete. Moreover, from a practical point of view, even if a
perfect entangled state has been generated in a laboratory we
cannot guarantee its entangled character after interaction with the
environment, due to unavoidable quantum noise (e.g. in long
distance quantum communication). Thus efficient detection of
entanglement is crucial for various quantum information tasks.

In 1964, John Bell showed that no local hidden-variable theory can
reproduce all of the statistical predictions of quantum mechanics
\cite{Bell64}. This was developed further in the form of the
Clauser-Horne-Shimony-Holt inequality (CHSH-Bell inequality) for
experimentally testing nonlocal quantum correlation between two
separated entangled particles \cite{CHSH}, and was first
demonstrated experimentally by Aspect \textit{et al }\cite{Aspect}.
In general, however, we have to take into account all possible
settings for all the local observables that appear in the Bell
inequality to test its violation. This is not efficient
experimentally if we have no prior knowledge of a given
state. Also, there are some entangled states which do not violate
the Bell inequality, such as some of the Werner states
\cite{werner89}. In this case an alternative method is to use
quantum state tomography \cite{tomog1,tomog2} to obtain the
complete density matrix for a quantum state and then apply certain
known sufficient or necessary entanglement criteria. (For recent
good reviews we refer to \cite{lbck00,3hreview,terhal01} and
references therein.) Among these, the Peres-Horodecki criterion
\cite{peres,3hPLA223}, the recent realignment criterion
\cite{ru02,Chen02,chenPLA02} and its multipartite generalization
``the generalized partial transposition criterion''
\cite{chenPLA02} are three strong operationally-friendly
entanglement criteria which can fully recognize entanglement in
$2\times2$ and $2\times3$ systems as well as distinguish most bound
entangled states (which are not distillable) in higher dimensions.
Moreover, we can use Wootters's elegant formula to calculate the
entanglement of formation for two qubits \cite{wo98}. The big
disadvantage of these methods is that we have to make a large
number of measurements ($4^{2}-1=15$ parameters for two qubits) to
determine the complete density matrix.

Recently, much effort has been devoted to finding ways to detect
the entanglement directly without having to measure the whole
density matrix. For pure states, a possible optimal strategy is
given in \cite{sanhue00}. Horodecki and Ekert proposed a method
based on structural physical approximations and collective
measurements that can be applied to mixed states \cite{Horo-Ekert}.
For 2 qubits this only needs 4 parameters to determine the degree
of entanglement \cite{horo0111082}, but in practice it requires the
construction of quantum gates and networks, which is not easy to
implement with present experimental technology. If, however, we are
given some \emph{prior} knowledge of a quantum state, a few local
measurements are enough to detect entanglement for two or three
qubits and certain bound entangled states \cite{guhne}. This has
been generalized \cite{pit-rubin} to higher dimensions and some
families of $n$ qubits by making use of the geometrical character
of entanglement witnesses \cite{ter,lew00,lew01,sanpera01}. For
depolarized states of bipartite systems in arbitrary dimensions
there is another scheme which only requires three local
measurements \cite{amp02}. However, these procedures all require
some prior knowledge of the quantum state and are only efficient
for special classes of states, so they are of limited use in
realistic quantum information processing.

In this Letter we develop a new method of entanglement detection
for two qubits which requires only a few local measurements and
\emph{no prior} knowledge of the quantum state. We also show its
relationship with the CHSH-Bell inequality and the entanglement
measure in terms of the concurrence \cite{wo98}. With the
experimental technology currently available it should not be too
difficult to implement the test.

In Hilbert space
$\mathcal{H}=\mathcal{C}^{2}\otimes\mathcal{C}^{2}$ it is well
known that we can represent the density matrix of any two qubits by
the bases in terms of the Kronecker product of the Pauli matrices,
as follows:
\begin{equation}
\rho=\frac{1}{4}\sum_{i,j=0}^{3}R_{ij}\sigma_{i}\otimes\sigma_{j},
\end{equation}
where $\sigma_{0}$ is the identity operator and $\sigma_{1,2,3}$
are the standard Pauli matrices. Here $R_{ij}$ is real and can be
calculated as $R_{ij} =Tr(\rho\sigma_{i}\otimes\sigma_{j})$ since
$Tr(\sigma_{i} \sigma_{j})=2\delta_{ij}$ and $\delta_{ij}$ is the
Kronecker delta symbol. Thus $R$ is a $4\times4$ real matrix and a
representation for the original density matrix $\rho$. For
convenience, we denote the $3\times3$ sub-matrix $[R_{ij}]$
$(i,j=1,2,3)$ by $T_{\rho}$.

We shall now derive a practical detection method by using only 9
expectation values of the local observables
$\sigma_{i}\otimes\sigma_{j}$.

\vspace*{5pt} \noindent\textbf{Theorem 1:} \label{theorem}
\emph{For any separable state of two qubits, the trace norm
$\left\| T_{\rho}\right\|$ of $T_{\rho}$, which is the sum of all
the singular values $s_{i}$ of $T_{\rho}$, is less than or equal
to 1, that is, $\left\| T_{\rho}\right\|\equiv\sum_{i=1}^{3}s_{i}
(T_{\rho})\leq1$, while the state is entangled if $\left\| T_{\rho}\right\| >1$.}

\vspace*{5pt} \noindent\textbf{Proof:} A separable quantum state
is a state which cannot be prepared locally and in which there is
no quantum correlation. Mathematically, this means that the
density matrix $\rho$ can be decomposed into an ensemble of
product states:
\begin{equation}
\rho=\sum_{i}p_{i}\rho_{i}^{A}\otimes\rho_{i}^{B}
\end{equation}
where $\rho_{i}^{A}=\left| \psi_{i}\right\rangle _{A}\left\langle
\psi _{i}\right| $, $\rho_{i}^{B}=\left| \phi_{i}\right\rangle
_{B}\left\langle \phi_{i}\right| $, $\sum_{i}p_{i}=1$ and $\left|
\psi_{i}\right\rangle _{A} $, $\left| \phi_{i}\right\rangle _{B}$
are normalized pure states of the subsystems $A$ and $B$,
respectively \cite{werner89}. It should be noted that
$\rho_{i}^{A,B}$ can be characterized by $3$-dimensional real
vectors (Bloch vectors) $(\lambda_{1}^{i},\lambda_{2}^{i},\lambda_{3}^{i})$
and $(\eta _{1}^{i},\eta_{2}^{i},\eta_{3}^{i})$, respectively, as
\begin{equation}
\rho_{i}^{A}=\frac{1}{2}\sum_{k=0}^{3}\lambda_{k}^{i}\sigma_{k},\rho_{i}
^{B}=\frac{1}{2}\sum_{k=0}^{3}\eta_{k}^{i}\sigma_{k},
\end{equation}
where $\lambda_{0}^{i}=\eta_{0}^{i}=1.$ The conditions $(\lambda_{1}^{i}
)^{2}+(\lambda_{2}^{i})^{2}+(\lambda_{3}^{i})^{2}=(\eta_{1}^{i})^{2}+(\eta
_{2}^{i})^{2}+(\eta_{3}^{i})^{2}=1$ should be satisfied because the set of
pure states corresponds to the surface of the Bloch sphere. Thus
we have
\begin{equation}
\rho_{i}^{A}\otimes\rho_{i}^{B}=\frac{1}{4}\sum_{k=0}^{3}\lambda_{k}^{i}
\sigma_{k}\otimes\sum_{l=0}^{3}\eta_{l}^{i}\sigma_{l},
\end{equation}
and $R_{kl}^{i}=Tr((\rho_{i}^{A}\otimes\rho_{i}^{B})\sigma_{k}\otimes
\sigma_{l})=\lambda_{k}^{i}\eta_{l}^{i}$. It is obvious then that
$T_{\rho_{i}^{A}\otimes\rho_{i}^{B}}=(\lambda_{1}^{i},\lambda_{2}^{i},
\lambda_{3}^{i})^{t}(\eta_{1}^{i},\eta_{2}^{i},\eta_{3}^{i})$ where $t$
denotes the standard transposition. Furthermore, it is clear that
$\left\|T_{\rho _{i}^{A}\otimes\rho_{i}^{B}}\right\| =\left\|
(\lambda_{1}^{i},\lambda _{2}^{i},\lambda_{3}^{i})^{t}\right\| \times\left\|
(\eta_{1}^{i},\eta _{2}^{i},\eta_{3}^{i})\right\| =1.$ Therefore, $\left\|
T_{\rho}\right\| \leq\sum_{i}p_{i}\left\|
T_{\rho_{i}^{A}\otimes\rho_{i}^{B}}\right\| =1$ due to the convex property
of the trace norm.\hfill\rule{1ex}{1ex}

\vspace*{5pt} To detect entanglement we only need to measure the
expectation values of the 9 local observables
$\sigma_{i}\otimes\sigma_{j} $ to obtain $(T_{\rho })_{i,j}$, then
compare the trace norm of $T_{\rho}$ and $1$. One question is
immediate: is this test stronger or weaker than the standard Bell
inequality test? According to \cite{CHSH,3hPLA200,frankPRL02}, the
CHSH-Bell test can be formulated using the expectation value
$\left\langle \mathcal{B}\right\rangle$ of the Bell operator
\begin{equation}
\mathcal{B}=\sum_{ij=1}^{3}\big( a_{i}(c_{j}+d_{j})+b_{i}(c_{j}-d_{j})\big)
\sigma_{i}\otimes\sigma_{j},
\end{equation}
with $(\overrightarrow{a},\overrightarrow{b},\overrightarrow{c},
\overrightarrow{d})$ being any real unit vectors and $\sigma_{i}$
the Pauli matrices, so that $\left\langle \mathcal{B}\right\rangle
=Tr(\rho\mathcal{B})\leq2$ should be satisfied by any local
classical model. An important advance made by the Horodecki
family \cite{3hPLA200} gave a necessary and sufficient condition
for two qubits to violate the CHSH-Bell test: the inequality is
violated iff $s_{1}^{2}+s_{2}^{2}>1$, where $s_{1}$ and $s_{2}$ are
two of the maximal singular values of $T_{\rho}$. Applying this
result, we derive a close relationship between Theorem $1$ and the
Bell inequality test:

\vspace*{5pt} \noindent\textbf{Theorem 2:} \emph{Detection of
Theorem 1 is stronger than the Bell inequality, i.e. any
entanglement which can be detected by the Bell inequality can also
be detected by the condition $\left\| T_{\rho}\right\| >1.$}

\vspace*{5pt} \noindent\textbf{Proof:} For any state satisfying
$\left\| T_{\rho}\right\| \leq1$ we have $s_{1}+s_{2}+s_{3}\leq1$,
so $s_{1}^{2}+s_{2}^{2}\leq s_{1}^{2}+s_{2}^{2}+s_{3}^{2}\leq
s_{1}+s_{2}+s_{3}\leq1$ since $s_{i}\geq0$ and $s_{i}^{2}\leq
s_{i}$. Thus any state violating the Bell inequality satisfies
$s_{1}^{2}+s_{2}^{2}>1$ and gives $s_{1}+s_{2}+s_{3}>1$, which
cannot escape detection by Theorem $1$. \hfill\rule{1ex}{1ex}

\vspace*{5pt} Now we would like to know how entangled is a given
quantum state. For two qubits the degree of entanglement, in terms
of the \textit{entanglement of formation} \cite{be96}, can be
calculated by the elegant formula of Wootters
\cite{wo98}:
\begin{equation}
E_{f}(\rho)=h\left( \frac{1+\sqrt{1-C^{2}}}{2}\right),  \label{EoF}
\end{equation}
where $h(x)=-x\log{x}-(1-x)\log{(1-x)}$ and the \textit{concurrence}
$C=\max\left[ 0,\tau_{1}-(\tau_{2}+\tau_{3}+\tau_{4})\right] $ with
$\{\tau_{i}^{2}\}$ being the decreasingly ordered eigenvalues of $\rho
(\sigma_{2}\otimes\sigma_{2})\rho^{T}(\sigma_{2}\otimes\sigma_{2})$. Noting
that $E_{f}(\rho)$\textit{\ }is in fact a convex and monotone function with
respect to the concurrence $C$, we can use the concurrence for convenience
in the following. For any pure state in the standard basis:
\begin{equation}
\left| \psi\right\rangle =a\left| 00\right\rangle +b\left| 01\right\rangle
+c\left| 10\right\rangle +d\left| 11\right\rangle,  \label{purestate}
\end{equation}
where $a,b,c,d$ are complex numbers satisfying
$|a|^{2}+|b|^{2}+|c|^{2}
+|d|^{2}=1$, we have a simpler expression for the concurrence $C(\left|
\psi\right\rangle )=2|ad-bc|$. Now we can see that Theorem $1$ gives further
an estimation for the amount of entanglement in terms of the concurrence:

\vspace*{5pt} \noindent\textbf{Theorem 3:} \emph{For any pure state $\left|
\psi\right\rangle $, $\frac{\left\| T_{\left| \psi\right\rangle }\right\| -1
}{2}$ is equal to the concurrence $C(\left| \psi\right\rangle )$. For any
mixed state it gives a lower bound for the the concurrence $C(\rho)$, i.e.
$\frac{\left\| T_{\rho}\right\| -1}{2}\leq C(\rho)$.}

\vspace*{5pt} \noindent\textbf{Proof:} For the pure state
$\left|\psi \right\rangle $ of Eq. \ref{purestate}, we have
$(T_{\left|\psi\right\rangle
})_{ij}=Tr((a,b,c,d)^{t}(a^{\ast},b^{\ast},c^{\ast},d^{\ast})\sigma_{i}
\otimes\sigma_{j})$. It is straightforward to calculate the eigenvalues
$s_{1}^{2},s_{2}^{2},s_{3}^{2}$ of $T_{\left| \psi\right\rangle }T_{\left|
\psi\right\rangle }^{t}$ and to obtain $\left\| T_{\left| \psi\right\rangle
}\right\| =s_{1}+s_{2}+s_{3}=4|ad-bc|+1=2C(\left|
\psi\right\rangle )+1$. Thus we have $C(\left|
\psi\right\rangle )=\frac{\left\| T_{\left| \psi\right\rangle }\right\| -1}
{2}$. Suppose that for the mixed state $\rho$ we have the
decomposition which gives the concurrence $C(\rho)$. That is,
\begin{equation}
C(\rho)=\sum_{i}p_{i}C(\left| \psi\right\rangle _{i}),
\end{equation}
where $\rho=\sum_{i}p_{i}\left| \psi\right\rangle _{i}\left\langle
\psi\right| $ and $\sum_{i}p_{i}=1$. It is natural that we have $\left\|
T_{\rho}\right\| =\left\| \sum_{i}p_{i}T_{\left| \psi\right\rangle _{i}
}\right\| \leq\sum_{i}p_{i}\left\| T_{\left| \psi\right\rangle _{i}}
\right\| =\sum_{i}p_{i}(2C(\left| \psi\right\rangle _{i})+1)=2C(\rho)+1$ due
to the convexity of the trace norm. Therefore $\frac{\left\| T_{\rho
}\right\| -1}{2}$ leads to a lower bound for the concurrence $C(\rho )$.
\hfill\rule{1ex}{1ex}

\vspace*{5pt} In most practical applications we do have some prior
knowledge of the quantum state for a given system. For example, we
can generate a known perfectly entangled pure state $\left|
\psi\right\rangle $ in one place and send it by some classical or
quantum channel to another place where we wish to know its final
state on arrival. Due to interaction with noise in the environment,
the state $\left| \psi\right\rangle$ will evolve to a mixed state.
One typical example is after going through a depolarizing channel,
the state will
transform to:
\begin{equation}
\rho=p\left| \psi\right\rangle \left\langle \psi\right| +(1-p)\mathbb{I}/4.
\label{depstate}
\end{equation}
where $\mathbb{I}/4$ is the maximally mixed state and $p$ is a
constant $0\leq p\leq 1$ representing the degree of depolarization.
This class of states can now be determined by Theorem 1. We know
that any pure state $\left| \psi\right\rangle $ of Eq.
\ref{purestate} can evolve through a Schmidt decomposition to
$\left| \psi^{^{\prime}}\right\rangle =\lambda_{1}\left|
00\right\rangle +\lambda_{2}\left| 11\right\rangle $ after a
unitary transformation of the local bases. Without loss of
generality, we suppose that $\left| \psi\right\rangle =a\left|
00\right\rangle +b\left| 11\right\rangle $ where $a,b>0$ and
$a^{2}+b^{2}=1$, from which we obtain
\begin{equation}
T_{\rho}=\left(
\begin{array}{ccc}
2abp & 0 & 0 \\
0 & 2abp & 0 \\
0 & 0 & -p
\end{array}
\right) .
\label{Trho}
\end{equation}
Thus $\left\| T_{\rho}\right\| =2|2abp|+|-p|=4abp+p$. Theorem $1$
says that $4abp+p>1$ implies entanglement, i.e. $1-p-4abp<0.$
Noticing that $(1-p)/4-abp$ gives the minimal eigenvalue for
partial transposition of $\rho$ with respect to the first
subsystem, our test is surprisingly equivalent to the
Peres-Horodecki criterion and can completely identify this class of
states. We only need to make three measurements of the local
observables $\sigma_{i}\otimes \sigma_{i}$ $(i=1,2,3)$ (or even
just two since $(T_{\rho})_{11}=(T_{\rho})_{22}$).

From the above-mentioned operations we notice that $1-\left\|
T_{\rho}\right\|
=1-(T_{\rho})_{11}-(T_{\rho})_{22}+(T_{\rho})_{33}=Tr(\rho(\mathbb{I}
-\sigma_{1}\otimes\sigma_{1}-\sigma_{2}\otimes\sigma_{2}+\sigma_{3}
\otimes\sigma_{3}))$. This suggests that
$\mathcal{W}=\mathbb{I}-\sigma
_{1}\otimes\sigma_{1}-\sigma_{2}\otimes\sigma_{2}+\sigma_{3}\otimes\sigma_{3}$
may act as a witness operator to detect entanglement. An
entanglement witness $\mathcal{W}$ is a Hermitian operator (an
observable) which satisfies $Tr(\mathcal{W}\rho\geq 0$ for all
separable states. Thus a state $\rho$ is entangled if we have
$Tr(\mathcal{W}\rho)< 0$ \cite{ter,lew00,lew01,sanpera01}. In fact,
$\mathcal{W}$ can here be expressed as
\begin{equation}
\mathcal{W}=2\left(
\begin{array}{cccc}
1 & 0 & 0 & 0 \\
0 & 0 & -1 & 0 \\
0 & -1 & 0 & 0 \\
0 & 0 & 0 & 1
\end{array}
\right) ,
\end{equation}
and is indeed an optimal witness, the same as the one given in
\cite{guhne} up to a constant factor $4$. Thus Theorem $1$ is
strong enough to detect any degree of entanglement in the class of
states represented by Eq. \ref{depstate}. It recognizes
entanglement in a subtle way while only requiring less than $3$
measurements of local observables. We can also derive an optimal
witness operator from the construction of our test, as well as
recover the result of \cite{amp02} which involves three local
observable measurements. However, our test is even better in that
we need only use two local measurements when we consider the
structural character of expression \ref{Trho}.

For pure states, a possible optimal strategy to detect entanglement is to measure the
reduced density matrix, as shown in \cite{sanhue00}.
However, measuring all the 9 local observables by means of Theorem 1 is a
costly matter if we have no prior knowledge of the state.
Here we propose a better strategy which requires few local
operations and uses only 3 observables:

\vspace*{5pt} \noindent\textbf{Proposition 1:} \emph{For any pure
state
$\left| \psi\right\rangle $ of Eq. \ref{purestate}, we have $\left\| \left(
\begin{array}{ccc}
R_{01} & R_{02} & R_{03}
\end{array}
\right) \right\| =\left\| \left(
\begin{array}{ccc}
R_{10} & R_{20} & R_{30}
\end{array}
\right) \right\| =\sqrt{1-C^{2}(\left| \psi\right\rangle )}$.}

\vspace*{5pt} \noindent\textbf{Proof:} For the pure state $\left|
\psi \right\rangle $ of Eq. \ref{purestate} we have
$R_{0i}=Tr((a,b,c,d)^{t}
(a^{\ast},b^{\ast},c^{\ast},d^{\ast})\sigma_{0}\otimes\sigma_{i})$ and
$R_{i0}=Tr((a,b,c,d)^{t}(a^{\ast},b^{\ast},c^{\ast},d^{\ast})\sigma_{i}
\otimes\sigma_{0})$. It is straightforward to verify that $\left\| \left(
\begin{array}{ccc}
R_{01} & R_{02} & R_{03}
\end{array}
\right) \right\| =\left\| \left(
\begin{array}{ccc}
R_{10} & R_{20} & R_{30}
\end{array}
\right) \right\| =\sqrt{1-4|ad-bc|^{2}}=\sqrt{1-C^{2}(\left| \psi
\right\rangle )}$. \hfill\rule{1ex}{1ex}

\vspace*{5pt} Proposition $1$ provides a better detection method
than Theorem $1$ for pure states and only involves $3$ local
observables of $\sigma _{0}\otimes\sigma_{i}$ or
$\sigma_{i}\otimes\sigma_{0} $ $(i=1,2,3)$. After measurement, we
can calculate the exact amount of entanglement such as the
entanglement of formation in terms of the concurrence $C(\left|
\psi\right\rangle )$. This is in some degree similar to the scheme
to measure the whole reduced density matrix proposed in
\cite{sanhue00}.

The above two examples (the depolarized state and the pure state)
show that, with some prior knowledge, we can greatly reduce the
number of local observables required for detecting entanglement.
Using Theorem $1$ and Proposition $1$ is much more efficient than
reconstructing the density matrix through quantum tomography. In
terms of the factor $f\equiv $``number of parameters $\times$
number of copies'' defined in \cite{horo0111082}, our method is
parametrically superior $(f=9\times9=81)$ to the density matrix
reconstruction schemes $(f=15\times15=225)$ and is comparable with
the proposal in \cite{horo0111082} $(f=4\times20=80)$. Our scheme
is feasible with present mature experimental technology and needs
no prior knowledge of the state.

In any case our results lead the way to a new form of entanglement
detection with few local measurements. It is stronger
than the Bell-CHSH inequality test. If no prior knowledge is
available, there exists an infinite number of Bell-CHSH
inequalities which must be tested before its violation can be
strictly proved. This is not efficient for practical applications.
Theorem $1$ also suggests that a stronger Bell-like inequality
exists if we further add the contribution of the third singular
values $s_{3}$ of $T_{\rho}$ to the degree of entanglement in
Theorem $1$. We only need a certain many-setting Bell-type
inequality which is maximally violated iff $s_{1}^{2}
+s_{2}^{2}+s_{3}^{2}>1$ (compare the standard maximal violation iff
$s_{1} ^{2}+s_{2}^{2}>1$) or even a weaker condition. This is in
fact possible, as proved in Ref. \cite{ypcz0301030}, where a
certain $3$-setting Bell-like inequality is shown to be a
sufficient and necessary criterion for separability.

We are now closer to solving the problem of finding the minimal
measurement cost of detecting entanglement. Our method gives a
better practical test of entanglement without whole state
estimation. It recovers previous results for some special classes
of states (the depolarized and the pure states) and can be
implemented with feasible experimental technology. Though our test
is not sufficient to detect all the entangled states, it is
stronger than the Bell-CHSH inequality. It also gives a lower bound
for the concurrence and thus an estimate of the amount of
entanglement for a given state. We expect that a similar result
should exist for the case of higher dimensions in a bi-partite
system. We leave this as an interesting open problem for future
study.

\emph{Acknowledgements:} the authors would like to
thank O. G\"{u}hne for introducing their work, and
Shao-Ming Fei and Fan Heng for stimulating discussions. K.C.
is grateful to Guozhen Yang for his continuous
encouragement. This work was supported by the Chinese
Academy of Sciences, the National Program for Fundamental
Research, the National Natural Science Foundation of China
and the China Postdoctoral Science Foundation.

\end{document}